\documentclass[twoside,twocolumn,9pt]{article}
\usepackage{extsizes}
\usepackage[super,sort&compress,comma]{natbib} 
\usepackage[version=3]{mhchem}
\usepackage[left=1.5cm, right=1.5cm, top=1.785cm, bottom=2.0cm]{geometry}
\usepackage{balance}
\usepackage{mathptmx}
\usepackage{sectsty}
\usepackage{graphicx} 
\usepackage{lastpage}
\usepackage[format=plain,justification=justified,singlelinecheck=false,font={stretch=1.125,small,sf},labelfont=bf,labelsep=space]{caption}
\usepackage{float}
\usepackage{fancyhdr}
\usepackage{fnpos}
\usepackage[english]{babel}
\addto{\captionsenglish}{%
  
}
\usepackage{array}
\usepackage{droidsans}
\usepackage{charter}
\usepackage[T1]{fontenc}
\usepackage[usenames,dvipsnames]{xcolor}
\usepackage{setspace}
\usepackage[compact]{titlesec}
\usepackage{hyperref}
\definecolor{cream}{RGB}{222,217,201}

\usepackage{bm}
\usepackage{amsfonts}
\usepackage{amsmath}
\usepackage{amssymb}
\usepackage{booktabs}
\usepackage[]{algorithm}
\usepackage{algpseudocode}

\begin{document}

\pagestyle{fancy}
\thispagestyle{plain}
\fancypagestyle{plain}{
%%%HEADER%%%
\renewcommand{\headrulewidth}{0pt}
}
%%%END OF HEADER%%%

%%%PAGE SETUP - Please do not change any commands within this section%%%
\makeFNbottom
\makeatletter
\renewcommand\LARGE{\@setfontsize\LARGE{15pt}{17}}
\renewcommand\Large{\@setfontsize\Large{12pt}{14}}
\renewcommand\large{\@setfontsize\large{10pt}{12}}
\renewcommand\footnotesize{\@setfontsize\footnotesize{7pt}{10}}
\makeatother

\renewcommand{\thefootnote}{\fnsymbol{footnote}}
\renewcommand\footnoterule{\vspace*{1pt}% 
\color{cream}\hrule width 3.5in height 0.4pt \color{black}\vspace*{5pt}} 
\setcounter{secnumdepth}{5}

\makeatletter 
\renewcommand\@biblabel[1]{#1}            
\renewcommand\@makefntext[1]% 
{\noindent\makebox[0pt][r]{\@thefnmark\,}#1}
\makeatother 
\renewcommand{\figurename}{\small{Fig.}~}
\sectionfont{\sffamily\Large}
\subsectionfont{\normalsize}
\subsubsectionfont{\bf}
\setstretch{1.125} %In particular, please do not alter this line.
\setlength{\skip\footins}{0.8cm}
\setlength{\footnotesep}{0.25cm}
\setlength{\jot}{10pt}
\titlespacing*{\section}{0pt}{4pt}{4pt}
\titlespacing*{\subsection}{0pt}{15pt}{1pt}
%%%END OF PAGE SETUP%%%

%%%FOOTER%%%
\fancyfoot{}
\fancyfoot[LO,RE]{\vspace{-7.1pt}\includegraphics[height=9pt]{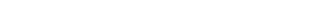}}
\fancyfoot[CO]{\vspace{-7.1pt}\hspace{11.9cm}\includegraphics{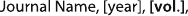}}
\fancyfoot[CE]{\vspace{-7.2pt}\hspace{-13.2cm}\includegraphics{head_foot/RF}}
\fancyfoot[RO]{\footnotesize{\sffamily{1--\pageref{LastPage} ~\textbar  \hspace{2pt}\thepage}}}
\fancyfoot[LE]{\footnotesize{\sffamily{\thepage~\textbar\hspace{4.65cm} 1--\pageref{LastPage}}}}
\fancyhead{}
\renewcommand{\headrulewidth}{0pt} 
\renewcommand{\footrulewidth}{0pt}
\setlength{\arrayrulewidth}{1pt}
\setlength{\columnsep}{6.5mm}
\setlength\bibsep{1pt}
%%%END OF FOOTER%%%

%%%FIGURE SETUP - please do not change any commands within this section%%%
\makeatletter 
\newlength{\figrulesep} 
\setlength{\figrulesep}{0.5\textfloatsep} 

\newcommand{\topfigrule}{\vspace*{-1pt}% 
\noindent{\color{cream}\rule[-\figrulesep]{\columnwidth}{1.5pt}} }

\newcommand{\botfigrule}{\vspace*{-2pt}% 
\noindent{\color{cream}\rule[\figrulesep]{\columnwidth}{1.5pt}} }

\newcommand{\dblfigrule}{\vspace*{-1pt}% 
\noindent{\color{cream}\rule[-\figrulesep]{\textwidth}{1.5pt}} }

\makeatother
%%%END OF FIGURE SETUP%%%

%%%TITLE, AUTHORS AND ABSTRACT%%%
\twocolumn[
  \begin{@twocolumnfalse}
{}\par
\vspace{1em}
\sffamily
\begin{tabular}{m{4.5cm} p{13.5cm} }
& \noindent\LARGE{\textbf{Bayesian estimation of optical constants using mixtures of Gaussian process experts}} \\%Article title goes here instead of the text "This is the title"
\vspace{0.3cm} & \vspace{0.3cm} \\

 & \noindent\large{Teemu Härkönen,$^{\ast}$\textit{$^{a}$} Hui Chen,\textit{$^{b}$} and Erik Vartiainen\textit{$^{c}$}} \\%Author names go here instead of "Full name", etc.

 & \noindent\normalsize{We propose modeling absorption spectrum measurements as mixtures of Gaussian process experts. This enables us to construct a flexible statistical model for interpolating and extrapolating measurements, facilitating statistical integration of Kramers-Kronig relations to estimate the whole complex refractive index. Additionally, we statistically model the anchoring points used in subtractive Kramers-Kronig relations to account for possible measurement errors of the anchor point. In addition to flexible statistical modeling, the mixtures of Gaussian process formulation enables automatic selection of measurement points to use for extrapolation. We apply the method to experimental absorption spectrum measurements of gallium arsenide, potassium chloride, and transparent wood.} \\

\end{tabular}

 \end{@twocolumnfalse} \vspace{0.6cm}

  ]
%%%END OF TITLE, AUTHORS AND ABSTRACT%%%

%%%FONT SETUP - please do not change any commands within this section
\renewcommand*\rmdefault{bch}\normalfont\upshape
\rmfamily
\section*{}
\vspace{-1cm}

%%%FOOTNOTES%%%
% 
\footnotetext{$^{*}$~\textit{Corresponding author, E-mail: teemu.h.harkonen@aalto.fi}}
\footnotetext{\textit{$^{a}$~Department of Electrical Engineering and Automation, Aalto University, Espoo, FI-02150, Finland}}
\footnotetext{\textit{$^{b}$~Department of Fibre and Polymer Technology, KTH Royal Institute of Technology, Stockholm, SE-10044, Sweden}}
\footnotetext{\textit{$^{c}$~Department of Computational Engineering, School of Engineering Sciences, LUT University, Yliopistonkatu 34, FI-53850, Lappeenranta, Finland}}

%%%MAIN TEXT%%%%
\section{Introduction}
\label{sec:introduction}
Kramers-Kronig relations are important tools in optics and materials science, as they connect the real and imaginary parts of a complex response function and are used in applications ranging from optical spectroscopy to materials characterization.\cite{Kramers:1927, Milton:1997, Jonscher:1977, Urquidi:1986, ODonnell:1981, Waters:2005}
In particular, they enable estimation of the real part of the complex refractive index from absorption measurements alone, provided that the absorption spectrum is known over the entire frequency range.

However, the numerical integration of the Kramers-Kronig relations suffers from truncated measurement data, resulting in poor estimates near the boundaries of the measurement domain and increased sensitivity to noise.
This limitation is realized when absorption measurements are available only over a finite frequency interval, which is almost always the case in practical experiments.
As a result, reliable extrapolation of the absorption spectrum outside the measurement range is crucial for accurate estimation of the refractive index.\cite{Aspnes1985_Ellipsometry}
The extrapolation requires two components, a mathematical model and a selection of points that are used to fit the mathematical model.
Typical approaches for the two are exponential decay and manual selection of points to use for the extrapolation specifically.
This is known to introduce errors in numerical Kramers-Kronig relations.\cite{Peiponen:1991, Marek:2002, Peiponen:2009, Mayerhofer:2022}
In particular, \citeauthor{Marek:2002} note that different extrapolation models result in significantly different results for the complex refractive index.
This presents an interesting computational problem of constructing an automatic, model-agnostic numerical integration of the Kramers-Kronig relations.

While using a parameterized model such as exponential decay, optimizing the model parameters, and performing extrapolation according to the optimized model
parameters is ubiquitous in science and engineering, a more flexible approach is to use a non-parametric model, and more specifically, a stochastic process.
A widely-used\cite{Liu:2020, Volker:2021, Bu:2022, Harkonen:2024, Otto:2024, Kuitunen:2025} example of such a model is the Gaussian process, which encodes prior information such as differentiability via its covariance function, or kernel.\citep{WilliamsRasmussen:2006}
Decaying random function samples can be generated by combining commonly used covariance functions, such as the squared exponential covariance function, with a linear covariance function, followed by exponentiation.
Formally, this covariance encoding allows us to assign a probability distribution on the possible extrapolation models instead of relying on a single explicit choice.

However, a single Gaussian process assumes identical noise levels and frequency behavior across the whole measurement range.
In practice, this means that, for example, sharp absorption features should follow a different Gaussian process compared to slowly varying regions of the absorption measurements.

General mixtures of experts\cite{ Masoudnia:2014, Gormley:2019} amend this by probabilistically partitioning a measurement space into distinct, independent regions.
The behavior of data and the underlying function in these regions are modeled with independent models.
These two mathematical mechanisms are called the gating network and experts, respectively.
Specifically, the gating network partitions the measurement data into independent sets to which the independent experts are fitted.
Assigning Gaussian processes as the individual experts results in a mixture of Gaussian process experts.
In addition to the flexible statistical modeling of the absorption measurements through the Gaussian process experts, the partitioning of the measurement space provides an automatic way of selecting which points are used for the extrapolation at the boundaries.
This constitutes an automatic and data-driven model for computing the complex refractive index through the Kramers-Kronig relations.
To fully incorporate uncertainty arising from the measurements, we statistically estimate all model parameters.
In particular, we use Bayesian inference.

Rather than assigning single best-fit values to unknown quantities, we treat the mixture model parameters and anchor point quantities as random variables.
The likelihood connects the measured attenuation spectrum to the parameters, while prior distributions encode weak regularity, such as parameter positivity, and plausible anchor point ranges.
Together, the likelihood and prior distribution define a posterior distribution for the parameters.
By constructing numerical parameter samples from the posterior distribution, we can sample the mixture of Gaussian process experts model for synthetic attenuation spectra.
These synthetic attenuation spectra can be propagated through the singly-subtractive Kramers-Kronig relation, yielding samples for the complex refractive index.
However, sampling the posterior distribution of the mixture model parameters is computationally challenging.
This has sparked multiple bespoke estimation approaches, such as maximum \textit{a posteriori} estimation for point estimates, variational inference, importance sampling, Gibbs sampling, and nested sequential Monte Carlo samplers. \cite{Tresp:2001, Yuan:2008, Zhang:2019, Gadd:2020, Harkonen:2025}
For our numerical example, we use the nested sequential Monte Carlo sampler due to its attractive unbiasedness and parallelizability properties.

Our main contribution is the mixture of Gaussian process experts modeling of absorbance measurements and statistical treatment of the anchor point in singly-subtractive Kramers-Kronig relations.
This allows automatic and robust interpolation and extrapolation of the imaginary part of the refractive index for statistical integration, resulting in posterior distributions for both the imaginary and real parts of the refractive index.
In particular, we use hierarchical Bayesian modeling at each step of our method to properly propagate uncertainty throughout the computation.

The remainder of this manuscript is organized as follows.
We discuss standard and singly-subtractive Kramers-Kronig relations in Section \ref{sec:kk-relations}.
In Sections \ref{sec:moe} and \ref{sec:statisticalKramersKronig}, we detail the mixture of Gaussian process experts model and the resulting statistical model for the complex refractive index.
We document our datasets in Section \ref{sec:data} and present our numerical results for the aforementioned datasets in Section \ref{sec:results}.
We conclude the study with discussion and contemplations in Section \ref{sec:conclusions}.

\section{Kramers-Kronig relations}
\label{sec:kk-relations}
For an analytic complex function $f(\hat\omega)$ of a complex frequency variable $\hat\omega = \omega + i\overline{\omega}$ in the upper half-plane, Kramers-Kronig relations describe the connection between the real and imaginary parts of $f(\omega)$ as
\begin{equation}
\begin{split}
    \text{Re}\left\{ f( \omega ) \right\} &= \frac{1}{\pi} \text{P} \int\limits_{-\infty}^\infty \frac{ \text{Im}\left\{ f( \omega' ) \right\} }{ \omega' - \omega } \text{d}\omega',\\
    \text{Im}\left\{ f( \omega ) \right\} &= -\frac{1}{\pi} \text{P} \int\limits_{-\infty}^\infty \frac{ \text{Re}\left\{ f( \omega' ) \right\} }{ \omega' - \omega } \text{d}\omega',
\end{split}
\label{eq:kkRelations}
\end{equation}
where $\text{P}$ denotes the Cauchy principal value.
For the complex refractive index $n ( \omega )$ with real refractive and imaginary attenuation components $\eta( \omega )$ and $\kappa( \omega )$
\begin{equation}
    n( \omega ) = \eta( \omega ) + i\kappa( \omega),
    \label{eq:complexRefractiveIndex}
\end{equation}
the corresponding Kramers-Kronig relations can be constructed as follows.
Assuming that high-energy photons experience no attenuation or refraction
\begin{equation}
    \lim_{\omega \rightarrow \infty} n( \omega ) = 1,
    \label{eq:complexRefractiveIndexLimit}
\end{equation}
together with an additional $\omega' + \omega$ factor included in Eq.~\eqref{eq:kkRelations}, we can algebraically impose parity on the integrals and obtain 
\begin{equation}
    \eta( \omega ) - 1 = \frac{2}{\pi} \text{P} \int\limits_0^\infty \frac{\omega' \kappa( \omega' ) }{ \omega'^2 - \omega^2 } \text{d}\omega',
    \label{eq:kkRelationEta}
\end{equation}
and
\begin{equation}
    \kappa( \omega ) = -\frac{2\omega}{\pi} \text{P} \int\limits_0^\infty \frac{ \eta( \omega' ) - 1 }{ \omega'^2 - \omega^2 } \text{d}\omega'.
    \label{eq:kkRelationKappa}
\end{equation}

The convergence and numerical stability of numerical integration of the relations in Eqs.~\eqref{eq:kkRelationEta} and \eqref{eq:kkRelationKappa} can be improved with singly subtractive Kramers-Kronig relations.
In particular extending Eq.~\eqref{eq:kkRelationEta} with singly subtractive Kramers-Kronig relations, \textit{a priori} known anchor or reference point $\left( \omega_\text{a}, \eta_\text{a} \right)$ is subtracted from Eq.~\eqref{eq:kkRelationEta}, resulting in
\begin{equation}
    \eta( \omega ) = \frac{2 \left(\omega^2 - \omega_\text{a}^2\right) }{\pi} \text{P} \int\limits_0^\infty \frac{\omega' \kappa( \omega' ) }{ \left(\omega'^2 - \omega^2\right) \left(\omega'^2 - \omega_\text{a}^2\right) } \text{d}\omega' + \eta_\text{a}.
    \label{eq:sskkRelationEta}
\end{equation}

Multiply subtractive Kramers-Kronig relations also exist for further convergence improvements.\cite{Palmer:1998}
However, we limit ourselves here to the singly subtractive relations for notational convenience.
For the exact formulation, see \citeauthor{Palmer:1998} or \citeauthor{Lucarini:2005:mskk}

Regardless of the attractive properties discussed above, the practical numerical computation suffers from a limited range of absorption spectrum measurements.
In the following Section, we present our mixture of Gaussian experts based approach for robust statistical computation of the singly subtractive Kramers-Kronig relations presented in Eq.~\eqref{eq:sskkRelationEta}.
\section{Mixtures of Gaussian process experts}
\label{sec:moe}
We model the logarithms of $N$ measured attenuation coefficients $\boldsymbol{\gamma} := \left( \gamma_1, \dots, \gamma_N \right)^\intercal := \left( \log\kappa( \omega_1 ), \dots, \log\kappa( \omega_N) \right)^\intercal $ at measurement frequencies $\boldsymbol{\omega} :=  \left( \omega_1, \dots, \omega_N\right)^\intercal$ as a mixture of experts (MoE).
An MoE partitions the measurement data into $K$ distinct sets of points
\begin{equation}
\begin{split}
     c_i \mid \left( \omega_i, \psi \right) & \sim \text{Categorical}\left( \pi( \omega_i \mid \psi) \right),
\end{split}
\end{equation}
where the allocation variable $c_i$ determines the partition into the $K$ experts based on the measurement location given $\psi$, the gating network parameters, with $ \pi( \omega_i \mid \psi) := ( \pi_1( \omega_i \mid \psi), \dots, \pi_K( \omega_i \mid \psi)) $ being the gating network which defines the categorical probability of each data point.
We construct the gating network using one-dimensional Gaussian kernels $ \left\{ \mathcal{K}_1( \omega_i \mid \psi_1), \dots, \mathcal{K}_K( \omega_i \mid \psi_K) \right\}$
\begin{equation}
    \pi_k(\omega_i \mid \psi ) = \frac{ \mathcal{K}_k( \omega_i \mid \psi_k)}{ \sum_{ k' = 1}^K \mathcal{K}_{k'}( \omega_i \mid \psi_{k'})},
    \label{eq:gatingnetwork}
\end{equation}
with
\begin{equation}
    \mathcal{K}_k( \omega_i \mid \psi_k) := v_k \mathcal{N}( \omega_i \mid \mu_k, \sigma_k^2),
    \label{eq:kernel}
\end{equation}
where $ v_k $ is the relative weight of the Gaussian kernel with mean $ \mu_k $ and variance $ \sigma_k^2 $, and $\psi_k = ( v_k, \mu_k, \sigma)^\intercal$. 

The gating network construction partitions the measurement data $ ( \boldsymbol{\omega}, \boldsymbol{\gamma})^\intercal $ into $K$ sets
denoted as $( \boldsymbol{\omega}_k, \boldsymbol{\gamma}_k) = \{ (\omega_i, \gamma_i) : c_i = k \}$.
Note that some of these partitioned sets can be empty.
Each of the partitioned datasets is modeled as an independent, zero-mean Gaussian process expert
\begin{equation}
    \boldsymbol{\gamma}_k \mid \left( \boldsymbol{\omega}_k, \theta_k \right) \sim \text{GP}\left( 0, \Sigma( \boldsymbol{\omega}_k, \boldsymbol{\omega}_k; \theta_k) + \sigma^2_{k,\epsilon}I \right),
    \label{eq:gpLikelihood}
\end{equation}
where $\Sigma( \boldsymbol{\omega}_k, \boldsymbol{\omega}_k; \theta_k)$ denotes the covariance function evaluated at the inputs $\boldsymbol{\omega}_k$, $\theta_k$ are are the parameters of covariance function including $\sigma^2_{k,\epsilon}$, the noise variance for the $k$th expert, and $I$ denotes the identity matrix.

For the covariance function, we use a sum of the squared exponential kernel and a linear kernel.
Therefore, each $ij$th element of the covariance matrix $\Sigma( \boldsymbol{\omega}_k, \boldsymbol{\omega}_k; \theta_k)$ is given by
\begin{equation}
\begin{split}
    \left[ \Sigma( \boldsymbol{\omega}_k, \boldsymbol{\omega}_k; \theta_k) \right]_{i,j} &= \sigma_{k,f}^2 \exp\left\{ -\frac{ (\omega_{i} - \omega_{j})^2 }{ l_{k}^2 } \right\}  + \boldsymbol{\omega}^\intercal\boldsymbol{\omega} + 1
\end{split}
\label{eq:covarianceKernel}
\end{equation}
where $\sigma_{k,f}^2$ and $l_k$ are the length scale and signal variance of the squared exponential kernel with  $\theta_k = (\sigma_{k,\varepsilon}, \sigma_{k,f}, l_{k} )$.
The linear kernel, $\boldsymbol{\omega}^\intercal\boldsymbol{\omega} + 1$ can include additional parameters.\cite{WilliamsRasmussen:2006}
However, in our particular application, additional linear kernel parameters were not identifiable.
In other words, practically any combination of the parameters caused a similar linear trend for the expert models.
Therefore, we opted to use the presented fixed form of the linear kernel.

The log-likelihood $\log\mathcal{L}\left( \boldsymbol{\gamma} \mid \boldsymbol{\omega}_k, \theta_k \right)$ of a single partitioned dataset $ \left( \boldsymbol{\omega}_k, \boldsymbol{\gamma}_k \right)^\intercal$ is given by
\begin{equation}
\begin{split}
    \log\mathcal{L}\left( \boldsymbol{\gamma}_k \mid \boldsymbol{\omega}_k, \theta_k \right) = &-\frac{1}{2} \boldsymbol{\gamma}_k^T \Sigma( \boldsymbol{\omega}_k, \boldsymbol{\omega}_k; \theta_k)^{-1} \boldsymbol{\gamma}_k \\ &-\frac{1}{2} \log \left\vert \Sigma( \boldsymbol{\omega}_k, \boldsymbol{\omega}_k; \theta_k) \right\vert - \frac{N_k}{2} \log 2\pi,
\end{split}
\end{equation}
where $\left\vert \Sigma( \boldsymbol{\omega}_k, \boldsymbol{\omega}_k; \theta_k) \right\vert$ is the determinant of the covariance matrix and $N_k$ is the number of data points associated with the $k$th expert.
As the $K$ experts are modeled as independent, the likelihood of the whole dataset given all expert parameters is
\begin{equation}
    \mathcal{L}( \boldsymbol{\gamma} \mid C, \boldsymbol{\omega}, \theta) = \prod\limits_{k = 1}^K \mathcal{L}\left( \boldsymbol{\gamma}_k \mid \boldsymbol{\omega}_k, \theta_k \right),
\end{equation}
where $C = ( c_1, \dots, c_N)^\intercal$ denotes a vector of allocation variables and $\theta = ( \theta_1, ..., \theta_K)^T$ is the combined vector of all the expert parameters.
For a detailed discussion of Gaussian processes, please see \citeauthor{WilliamsRasmussen:2006}.

In step with Bayesian methodology, we assign prior distributions for the gating network and expert parameters.
Typically, the prior distributions are described as incorporating \textit{a priori} information on the parameters.
However, interpreting them as a form of explicit regularization is a practical way of justifying their inclusion. 
We use identical prior distributions as used by \citeauthor{Harkonen:2025}, namely, half-normal distributions for the standard deviation parameters and expert length scales, a Dirichlet distribution with a concentration parameter of $0.5$, and normal distributions for the gating network means.
We denote the joint prior distribution of the gating network and Gaussian process expert parameters as $\pi_0( \psi )$ and $\pi_0( \theta )$, respectively.  
We detail the used prior distributions in Table \ref{table:priors}.

Compiling the above results in the following statistical model for the mixture of Gaussian process experts for the log-attenuation measurements
\begin{equation}
\begin{split}
    \boldsymbol{\gamma}_k \mid \left( \boldsymbol{\omega}_k, \theta_k \right) &\sim \text{GP}\left( 0, \Sigma( \boldsymbol{\omega}_k, \boldsymbol{\omega}_k; \theta_k) \right),\\
    c_i \mid \left( \omega_i, \psi \right) & \sim \text{Categorical}( \pi(\omega_i \mid \psi) ), \\
    \psi &\sim \pi_{0}(\psi), \\
    \theta &\sim \pi_0(\theta),
\end{split}
\end{equation}
which results in the following posterior distribution for the partition, together with the gating network and expert parameters
\begin{equation}
    \pi( C, \psi, \theta \mid \boldsymbol{\omega}, \boldsymbol{\gamma}) \propto \mathcal{L}( \boldsymbol{\gamma} \mid \boldsymbol{\omega}, C, \theta) \pi_0( C \mid \boldsymbol{\omega}, \psi) \pi_0( \psi ) \pi_0( \theta ),
    \label{eq:moeParameterPosteriorDistribution}
\end{equation}
where
\begin{equation}
    \pi_0( C \mid \boldsymbol{\omega}, \psi) \pi_0( \psi ) = \prod_{i=1}^N p(c_i \mid \omega_i, \psi) \prod_{k = 1}^K \pi_0( \psi_k )
    \label{eq:gatingNetworkPartitionPrior}
\end{equation}
encourages partitions that follow the gating network.
As the gating network partitions probabilistically, sampled partitions have different associated probabilities.
Eq.~\eqref{eq:gatingNetworkPartitionPrior} quantifies this, resulting in a regularizing effect on possible partitions.

The posterior distribution defined in Eq.~\eqref{eq:moeParameterPosteriorDistribution} is analytically intractable and is accessible only through numerical sampling or estimation.
We use a nested sequential Monte Carlo sampler specifically designed for sampling the posterior distribution presented in Eq.~\eqref{eq:moeParameterPosteriorDistribution}.
Due to the technicality of the nested sequential Monte Carlo sampler, we refer the reader to \citeauthor{Harkonen:2025} for a detailed discussion on the properties and implementation of the approach.

The sampler provides us with samples from the posterior distribution in Eq.~\eqref{eq:moeParameterPosteriorDistribution}.
In particular, the samples for the partition $C$, which define the datasets $( \boldsymbol{\omega}_k, \boldsymbol{\gamma}_k)^\intercal $, provide automated selection of data points to be used for extrapolation at the measurement boundaries.
The partitioned datasets $ ( \boldsymbol{\omega}_k, \boldsymbol{\gamma}_k)^\intercal $ together with the estimated gating network and expert parameters define the behavior of the attenuation at arbitrary frequency locations, allowing for flexible statistical interpolation, extrapolation, and ultimately integration of the attenuation measurements.
We detail this construction in the following Section.
\begin{table}
\centering
\begin{tabular}{c c}
    \hline
    Parameter & Prior distribution \\
    \hline
    $\mu_{k}$ & $\mathcal{N} \left( G_k, \left( \frac{ 0.25 }{ K + 1} \right)^2 \right)$ \\
    $\sigma_{k}$ & $\mathcal{N}_{\frac{1}{2}}\left(0, \left( \frac{ 0.25 }{ K + 1}\right)^2 \right)$ \\
    $\left( v_1, \dots, v_K \right)$ & $\rm{Dir}( 0.5, \dots, 0.5)$\\
    $\sigma_{k,\varepsilon}$ & $\mathcal{N}_{\frac{1}{2}} \left(0, \left(0.25\Delta\gamma\right)^2 \right)$ \\
    $\sigma_{k,f}$ & $\mathcal{N}_{\frac{1}{2}} \left(0, \left(0.25\Delta\gamma\right)^2 \right)$ \\
    $l_{k}$ & $\mathcal{N}_{\frac{1}{2}} \left(0, \left(0.50\Delta \omega\right)^2 \right)$ \\
    \hline
\end{tabular}
\caption{ Prior distributions for the mixture of Gaussian process experts model. The prior means for the gating networks' parameters $\mu_k$ are constructed using a linearly-spaced grid using spacing $1/K$, with $G_{k}= (G_{1},\ldots, G_{k})$ denoting the $k$th grid point. The prior variances for the variance and length scale parameters are based on data ranges, $\Delta\gamma = \max \boldsymbol{\gamma} - \min \boldsymbol{\gamma} $ and $ \Delta\omega = \max \boldsymbol{\omega} - \min \boldsymbol{\omega} $. Normal, half-normal, and Dirichlet distribution are denoted by $\mathcal{N}( \cdot, \cdot)$, $\mathcal{N}_{\frac{1}{2}}( \cdot, \cdot)$, and $\rm{Dir}( 0.5, \dots, 0.5)$, respectively. The Dirichlet distribution with a concentration parameter of $0.5$ corresponds to a non-informative prior distribution for the gating network weights.}
\label{table:priors}
\end{table}
\section{Statistical integration of Kramers-Kronig relations}
\label{sec:statisticalKramersKronig}
\begin{figure}
    \centering
    \includegraphics[width = \linewidth]{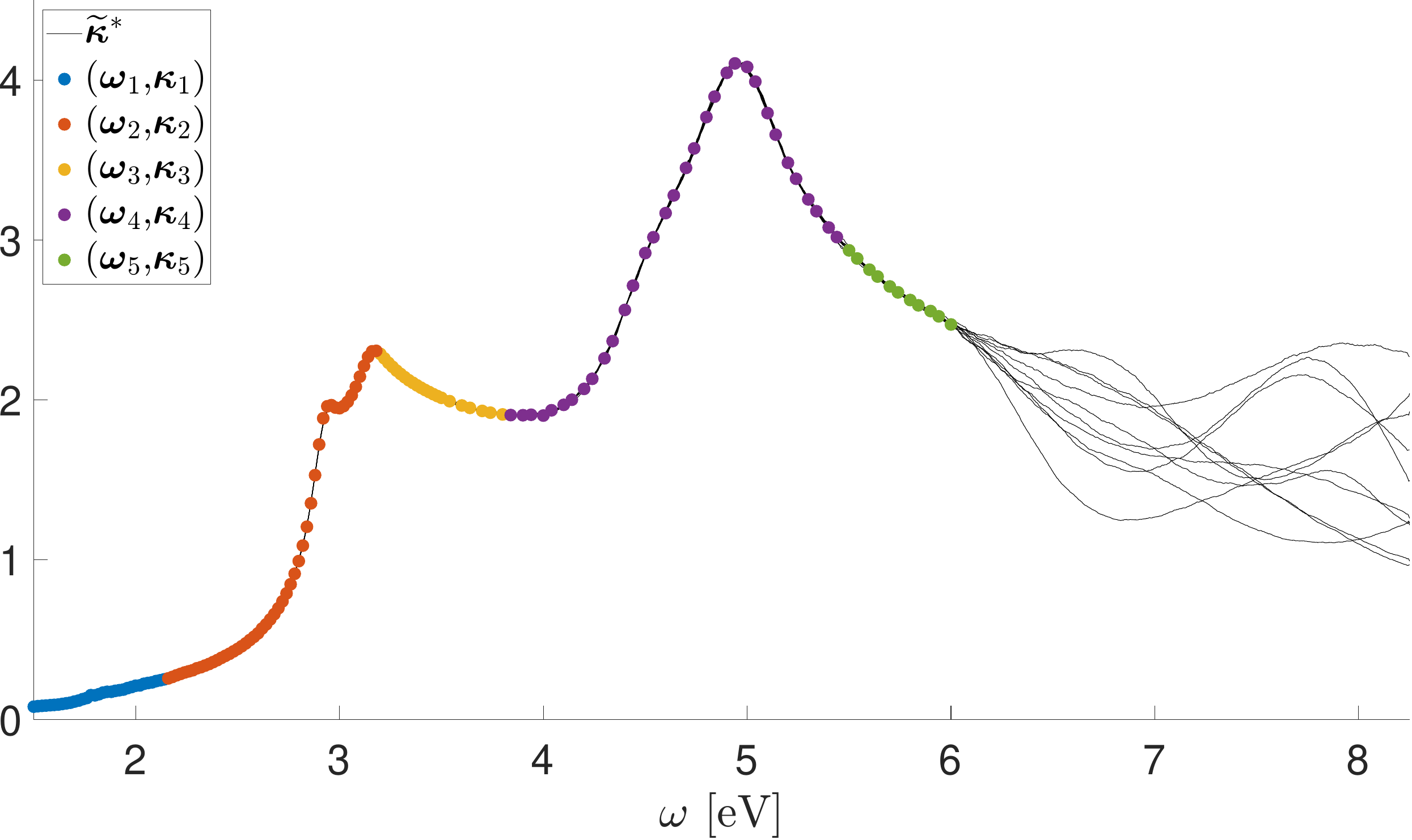}
    \caption{An example partition with $K = 5$ experts and 10 realizations drawn from the mixture of Gaussian process experts model estimated with nested sequential Monte Carlo sampling denoted by $(\boldsymbol{\omega}_k, \boldsymbol{\kappa}_k)$ and $\widetilde{\boldsymbol{\kappa}}^*$, respectively. Each realization can be used to compute a corresponding real part of the complex refractive index through Kramers-Kronig relations. The realizations closely follow the low-noise measurements. Outside the measurement range, $\omega \geq 6\text{ eV}$, the realizations show varied, slowly-decaying behavior encoded by the covariance function of the Gaussian process experts. The extrapolation is also performed for $\omega \leq 1.5\text{ eV}$, however, this part of the realizations is left out for visual clarity.}
    \label{im:examplePartitionRealization}
\end{figure}

Given the mixture of Gaussian process experts model parameter posterior distribution in Eq.~\eqref{eq:moeParameterPosteriorDistribution}, we can draw random samples, or realizations, for the attenuation coefficient $\kappa( \boldsymbol{\omega}^* )$ at arbitrary locations $\boldsymbol{\omega}^* = \left( \omega_1^*, \dots, \omega_{K^*}^* \right)^\intercal$.
The arbitrary locations allow for the aforementioned interpolation and extrapolation of the attenuation measurements.
The realizations are constructed via the analytical mean and covariance expressions for a single Gaussian process, which are then combined through spatial weighing according to the gating network.
Given $\theta_k$, the Gaussian process predictive mean $\boldsymbol{\mu}^* := \mu( \boldsymbol{\omega}^* )$ and covariance $ \Sigma^* := \Sigma^*( \boldsymbol{\omega}^*, \boldsymbol{\omega}^*, \theta_k)$ are given by
\begin{equation}
    \boldsymbol{\mu}_k^* = \Sigma( \boldsymbol{\omega}^*, \boldsymbol{\omega}; \theta_k) \Sigma( \boldsymbol{\omega}, \boldsymbol{\omega}; \theta_k)^{-1} \boldsymbol{\gamma}_k,
    \label{eq:gpPredictiveMean}
\end{equation}
and
\begin{equation}
    \boldsymbol{\Sigma}^* = \Sigma( \boldsymbol{\omega}^*, \boldsymbol{\omega}^*; \theta_k) - \Sigma( \boldsymbol{\omega}^*, \boldsymbol{\omega}; \theta_k) \Sigma( \boldsymbol{\omega}, \boldsymbol{\omega}; \theta_k)^{-1} \Sigma( \boldsymbol{\omega}^*, \boldsymbol{\omega}; \theta_k)^\intercal.
    \label{eq:gpPredictiveMean}
\end{equation}
By computing the Cholesky decomposition, $L$, of $ \boldsymbol{\Sigma}^* $, we can generate a random realization $\widetilde{ \boldsymbol{\gamma} }_k^* = \left( \widetilde{\gamma}_k^*( \omega_1^*), \dots, \widetilde{\gamma}_k^*( \omega_{K^*}^*) \right)^\intercal$ with
\begin{equation}
    \widetilde{ \boldsymbol{\gamma} }_k^* = \boldsymbol{\mu}_k^* + L\boldsymbol{\varepsilon},
\end{equation}
where $ \boldsymbol{\varepsilon} = \left( \varepsilon_1, \dots, \varepsilon_{N^*}\right)^\intercal $ is a vector of standard normal variables, $ \varepsilon_i \sim \mathcal{N}( 0, 1)$.
Repeating the process for all experts yields $K$ independent realizations which are combined into a single realization $\widetilde{ \boldsymbol{\gamma} }^* = \left( \widetilde{\gamma}^*( \omega_1^*), \dots, \widetilde{\gamma}^*( \omega_{K^*}^*) \right)^\intercal $ by
\begin{equation}
    \widetilde{\gamma}^*( \omega_i^*) = \sum\limits_{k = 1}^K \widetilde{ \gamma }_k^*( \omega_i^* ) \pi_k(\omega^*_i \mid \psi )
    \label{eq:moeGammaRealization}
\end{equation}
where the gating network weighs each realization according to the gating network relevance $\pi_k(\omega^*_i \mid \psi )$ at each location $ \omega_i^* $.
Finally, an attenuation realization is achieved through exponentiation of the log-attenuation, $ \widetilde{ \boldsymbol{\kappa} }^* = \exp\left\{ \widetilde{ \boldsymbol{\gamma} }^* \right\}$.
We present an example partition and realizations $\widetilde{\boldsymbol{\kappa}}^*$ in Figure \ref{im:examplePartitionRealization}.

\begin{figure}
    \centering
    \includegraphics[width = \linewidth]{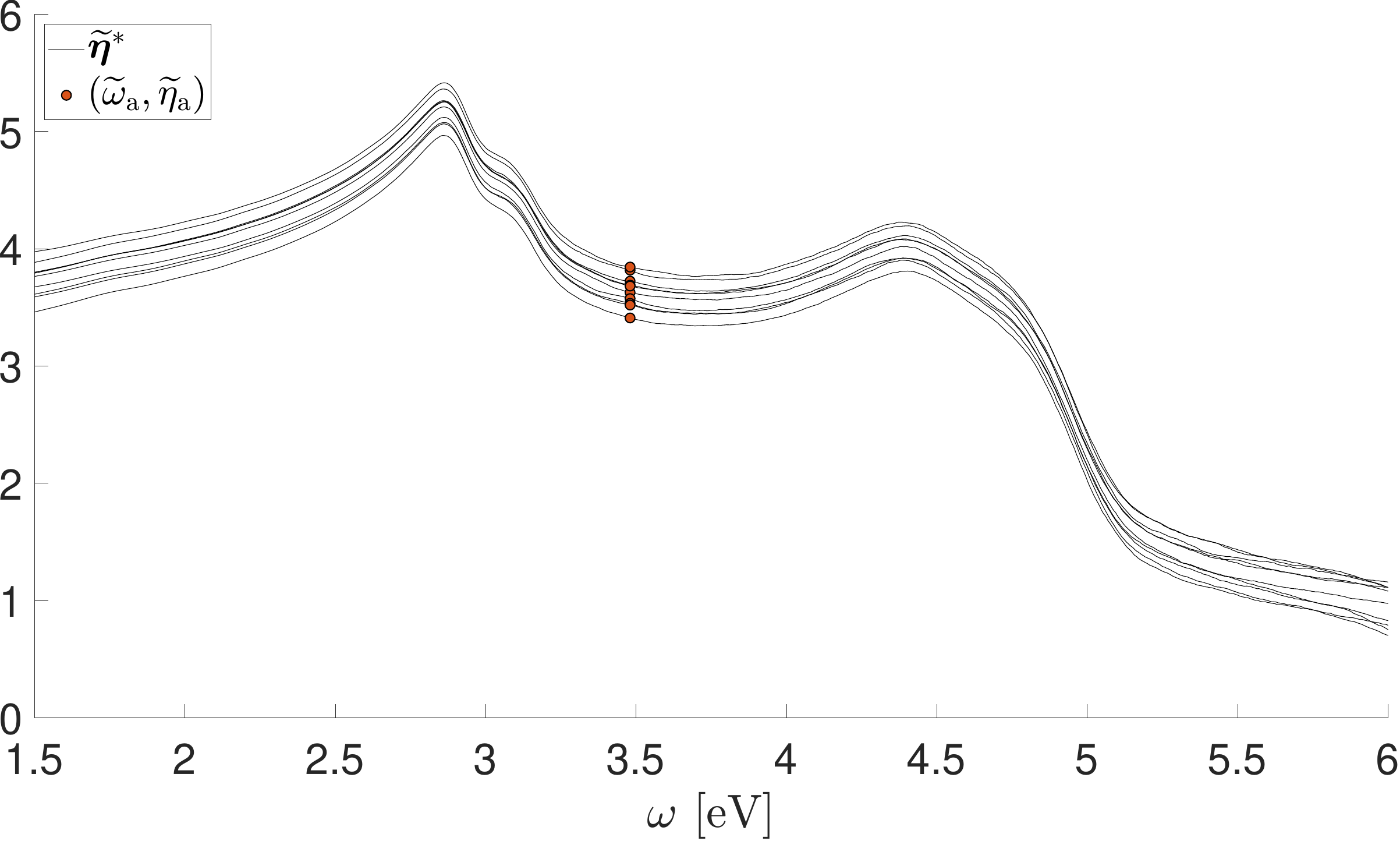}
    \caption{Example realizations $\widetilde{\boldsymbol{\eta}}^*$ corresponding to realizations $\widetilde{ \boldsymbol{\kappa} }^*$ in Figure \ref{im:examplePartitionRealization} and anchor points $\left( \widetilde{\omega}_\text{a}, \widetilde{\eta}_\text{a} \right)$ sampled from $\pi_0( \omega_\text{a}, \eta_\text{a} ) $. With a large ensemble of realizations, we can construct a numerical approximation for the complex refractive index distribution, $\pi( \boldsymbol{n}^* \mid \boldsymbol{\omega}^*, \boldsymbol{\omega}, \boldsymbol{\kappa} )$.}
    \label{im:exampleRealizationsEta}
\end{figure}

Given a realization $\widetilde{\boldsymbol{\kappa}}^*$ and an anchor point $\left( \omega_\text{a}, \eta_\text{a} \right)$, we can numerically integrate Eq.~\eqref{eq:sskkRelationEta} to obtain a sample for the refractive index, $\widetilde{\boldsymbol{\eta}}^* = \left( \widetilde{\eta}^*( \omega_1^*), \dots, \widetilde{\eta}^*( \omega_{K^*}^*) \right)^\intercal $.
We denote this numerical integration of the singly-subtractive Kramers-Kronig relations with $ I_\text{SSKK}\left( \boldsymbol{\omega}^*, \widetilde{\boldsymbol{\kappa}}^*, \omega_\text{a}, \eta_\text{a} \right) $.
As a last statistical modeling step, we assign a probability distribution for both the location and magnitude of the anchor point, $\pi_0( \omega_\text{0} )$ and $\pi_0( \eta_\text{a} )$ which we jointly denote by $\pi_0( \omega_\text{0}, \eta_\text{a}) = \pi_0( \omega_\text{0} )\pi_0( \eta_\text{a} ) $.
These anchor point distributions can be interpreted as modeling uncertainties arising from measurement laser frequency fluctuations or inaccuracies for the location, and noisy refractive index measurements or unknown bulk material properties for $\eta$. 
We can again compile the above into the following statistical model
\begin{equation}
\begin{split}
    \boldsymbol{\eta}^* &= I_\text{SSKK}\left( \boldsymbol{\omega}^*, \exp\{ \boldsymbol{\gamma}^* \}, \omega_\text{a}, \eta_\text{a} \right),\\
    \boldsymbol{\gamma}^* \mid \left( \boldsymbol{\omega}^*, C, \psi, \theta \right) &\sim \text{MoE}( \boldsymbol{\omega}^*; C, \psi, \theta )\\
    \left( C, \psi, \theta \right) &\sim \pi( C, \psi, \theta \mid \boldsymbol{\omega}, \boldsymbol{\gamma})
\end{split}
\label{eq:sskkStatisticalModel}
\end{equation}
where $ \text{MoE}( \boldsymbol{\omega}^*; C, \psi, \theta ) $ denotes the mixture of Gaussian process experts model, which we can sample as defined in Eqs.~\eqref{eq:gpPredictiveMean}--\eqref{eq:moeGammaRealization}.
In practice, given samples for $ \left( C, \psi, \theta \right) $, we can repeatedly sample realizations $\widetilde{\boldsymbol{\kappa}}^*$ and $\widetilde{\boldsymbol{\eta}}^*$ for each $ \left( C, \psi, \theta \right) $ sample, generating an ensemble of realizations which characterize the whole complex refractive index.
In Figure \ref{im:exampleRealizationsEta}, we present example realizations $\widetilde{\boldsymbol{\eta}}^*$ corresponding to the attenuation realizations presented in Figure \ref{im:examplePartitionRealization}. 
Formally, this sampling is a practical implementation of the following marginalization over the gating network and expert parameters
\begin{equation}
    \pi( \boldsymbol{n}^* \mid \boldsymbol{\omega}^*, \boldsymbol{\omega}, \boldsymbol{\kappa} ) = \int_\Upsilon \pi( \boldsymbol{n}^* \mid \boldsymbol{\omega}^*, \boldsymbol{\omega}, \boldsymbol{\kappa}, C, \psi, \theta) \text{d}C\text{d}\psi\text{d}\theta,
    \label{eq:complexRefractiveIndexPosteriorDistribution}
\end{equation}
where $\pi( \boldsymbol{n}^* \mid \boldsymbol{\omega}^*, \boldsymbol{\omega}, \boldsymbol{\kappa} )$ denotes the posterior distribution for the complex refractive index $\boldsymbol{n}^* = \boldsymbol{\eta}^* + i\boldsymbol{\kappa}^*$ at $\boldsymbol{\omega}^*$ given the measurement data, $\pi( \boldsymbol{n}^* \mid \boldsymbol{\omega}^*, \boldsymbol{\omega}, \boldsymbol{\kappa}, C, \psi, \theta)$ is the posterior distribution for $\boldsymbol{n}^*$ defined by the statistical model in Eq.~\eqref{eq:sskkStatisticalModel} for $\boldsymbol{\eta}^*$ and $\boldsymbol{\kappa}^*$, and $ \left( C, \psi, \theta \right) \in \Upsilon $.
We present high-level pseudo-code for the computational process in Algorithm \ref{alg:gp-moe-kk}.
\begin{algorithm}
    \caption{Mixtures of Gaussian process experts for Kramers-Kronig relations}
    \label{alg:gp-moe-kk}
    \begin{algorithmic}
        \State Step 1: Generate samples from the posterior distribution $ \pi( C, \psi, \theta \mid \boldsymbol{\omega}, \boldsymbol{\gamma}) $ defined in Eq.~\eqref{eq:moeParameterPosteriorDistribution}.
        \State Step 2: For a random sample generated in Step 1, compute an ensemble of realizations $\widetilde{ \boldsymbol{\gamma} }^*$ according to Eq.\eqref{eq:moeGammaRealization}.
        \State Step 3: For each $\widetilde{ \boldsymbol{\gamma} }^*$, sample an anchor point from $\pi_0( \omega_\text{a}, \eta_\text{a} )$.
        \State Step 4: For each realization and anchor paint pair, compute the numerical singly-subtractive Kramers-Kronig relation $ I_\text{SSKK}\left( \boldsymbol{\omega}^*, \widetilde{\boldsymbol{\kappa}}^*, \omega_\text{a}, \eta_\text{a} \right) $.
    \end{algorithmic}
\end{algorithm}

\section{Data}
\label{sec:data}
The experimental datasets used in this study comprise complex refractive index data for gallium arsenide (GaAs) and potassium chloride (KCl), as well as absorption data for a transparent wood sample. For GaAs and KCl, reference refractive index $\nu(\omega)$ and extinction coefficient $\kappa(\omega)$ spectra were obtained from established literature values.
Specifically, the refractive index and extinction coefficient data for GaAs were obtained from the tabulated optical constants reported by \citeauthor{Palik:1985_GaAs}.
Likewise, the corresponding complex refractive index data for KCl were taken from the compilation of \citeauthor{Palik:1985_KCl} on KCl.
The experimental dataset for transparent wood consists of spectrally resolved optical measurements, including transmittance and reflectance spectra.
The transmittance and reflectance were measured using a UV-Vis spectrophotometer equipped with an integrating sphere over the wavelength range of $450-800$ nm.
The absorption spectra were derived from the measured transmittance and reflectance using standard relations and the Beer-Lambert law.
These datasets provide material responses across a broad spectral range, enabling us to validate the statistical Kramers–Kronig framework developed in this work.
\section{Numerical examples}
\label{sec:results}
We apply the proposed method to the three datasets described in the previous Section.
For the transparent wood dataset, we additionally present results for a case where the Rayleigh scattering background signal has been removed using a 4th-order polynomial.
For all cases, we present results for the real, refractive part of the complex refractive index, which have been computed using the mixture of Gaussian process experts model with and without extrapolation to showcase the improvement at the boundaries of the measurement data.

\begin{figure}
    \centering
    \includegraphics[width = \linewidth]{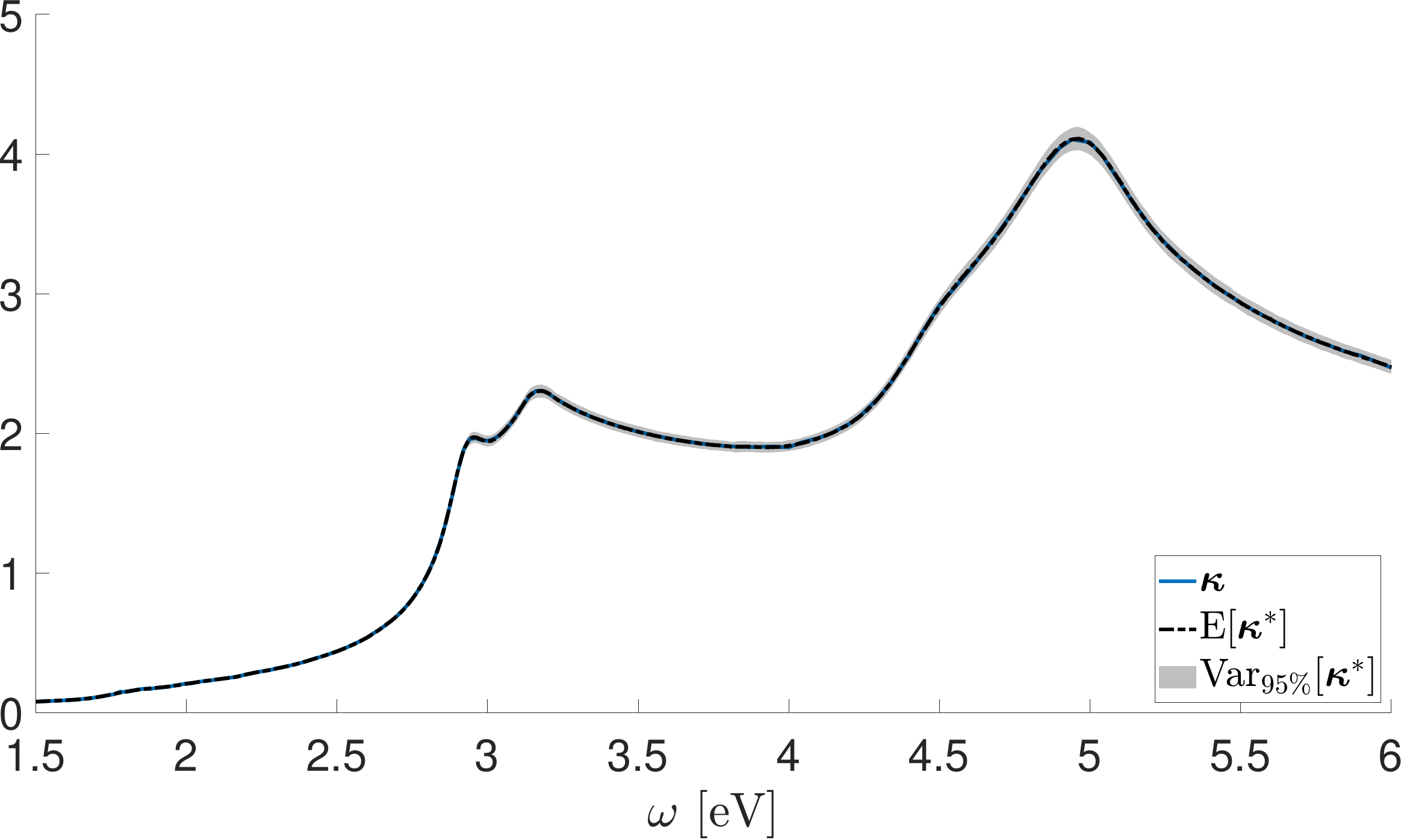}
    \includegraphics[width = \linewidth]{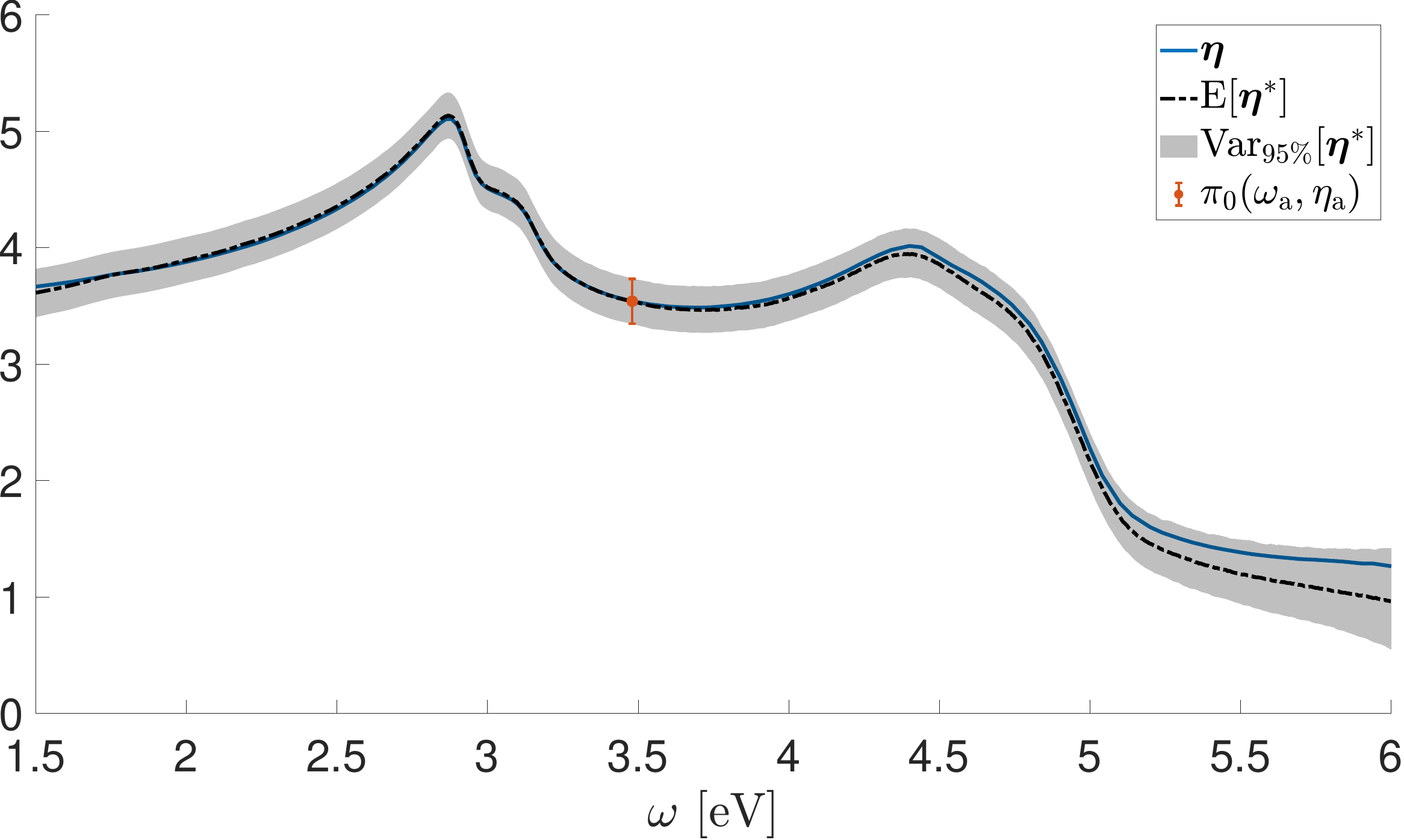}
    \includegraphics[width = \linewidth]{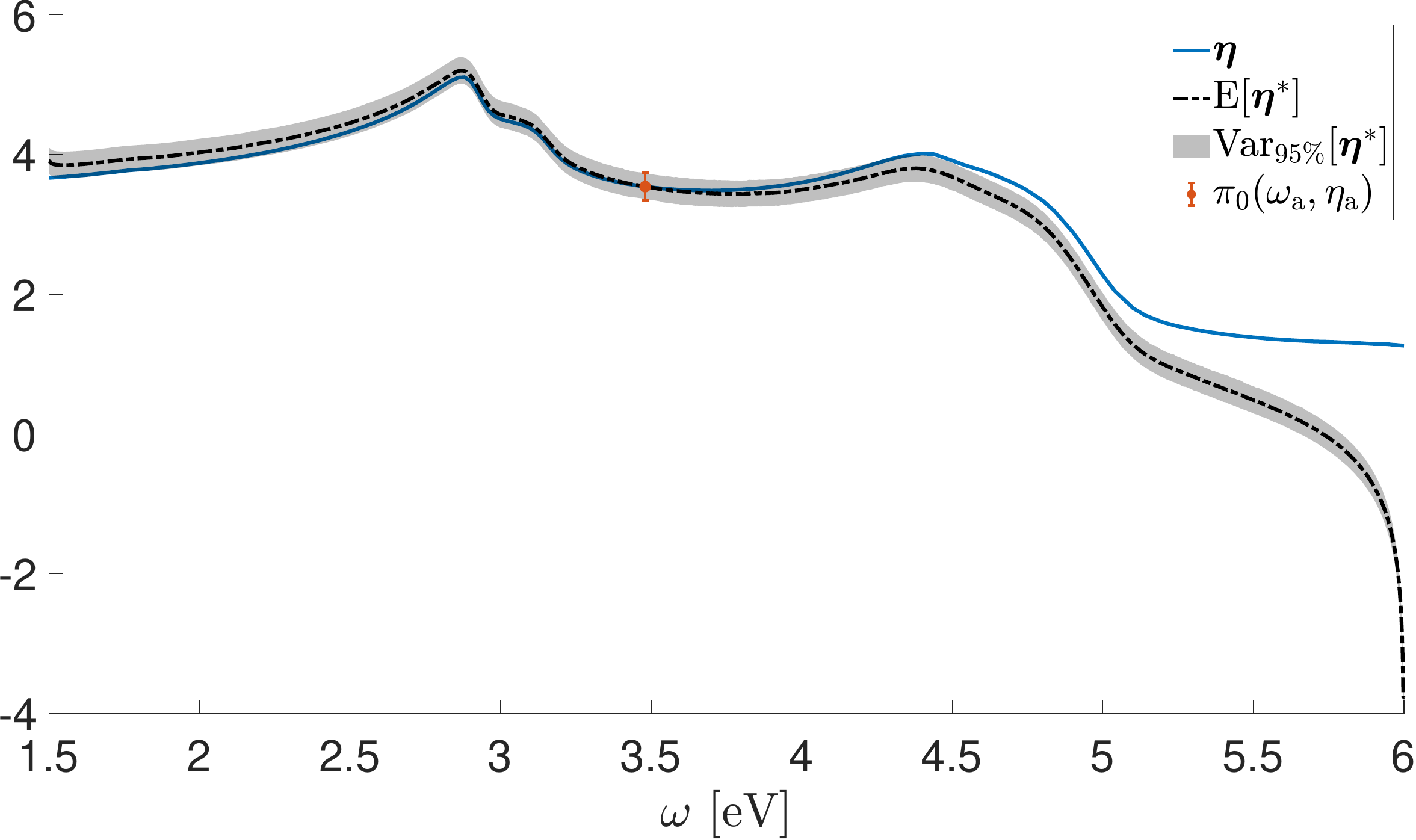}
    \caption{Results for the gallium arsenide dataset. At the top, attenuation measurement data $\boldsymbol{\kappa}$, together with estimated predictive mean and 95\% marginal predictive intervals denoted by $ \rm{E} \left[ \boldsymbol{\kappa}^* \right]$ and $ \rm{Var}_{95\%} \left[ \boldsymbol{\kappa}^* \right]$, respectively. In the middle and bottom, predictive means and 95\% marginal predictive intervals $ \rm{E} \left[ \boldsymbol{\eta}^* \right]$ and $ \rm{Var}_{95\%} \left[ \boldsymbol{\eta}^* \right]$ for the real, refractive component estimated with and without extrapolation, respectively. The reference measured refractive index is denoted by $\boldsymbol{\eta}$ and the 95\% intervals of the anchor point distribution $\pi_0( \omega_\text{a}, \eta_\text{a} )$ are illustrated in red.}
    \label{im:gaasResults}
\end{figure}

\begin{figure}
    \centering
    \includegraphics[width = \linewidth]{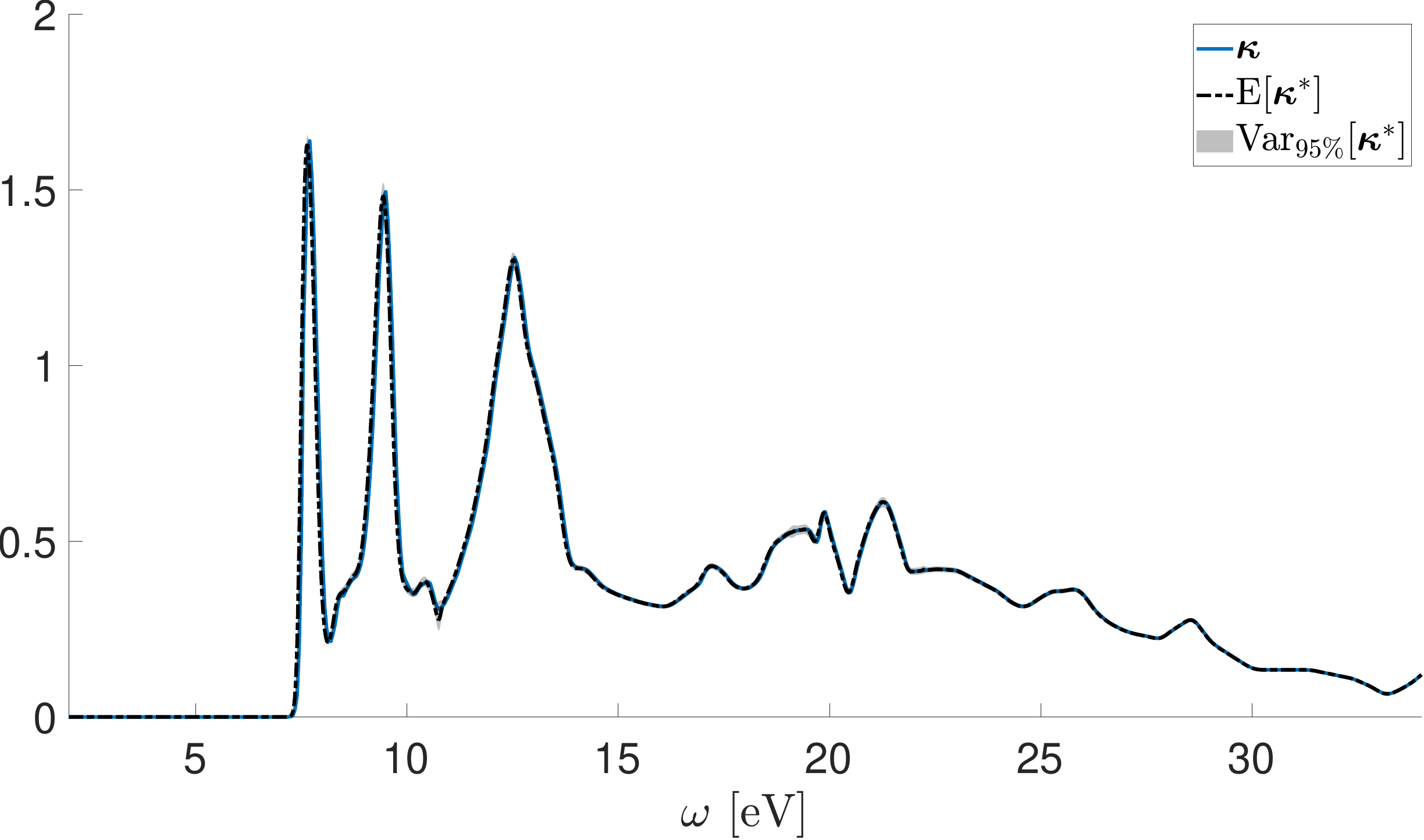}
    \includegraphics[width = \linewidth]{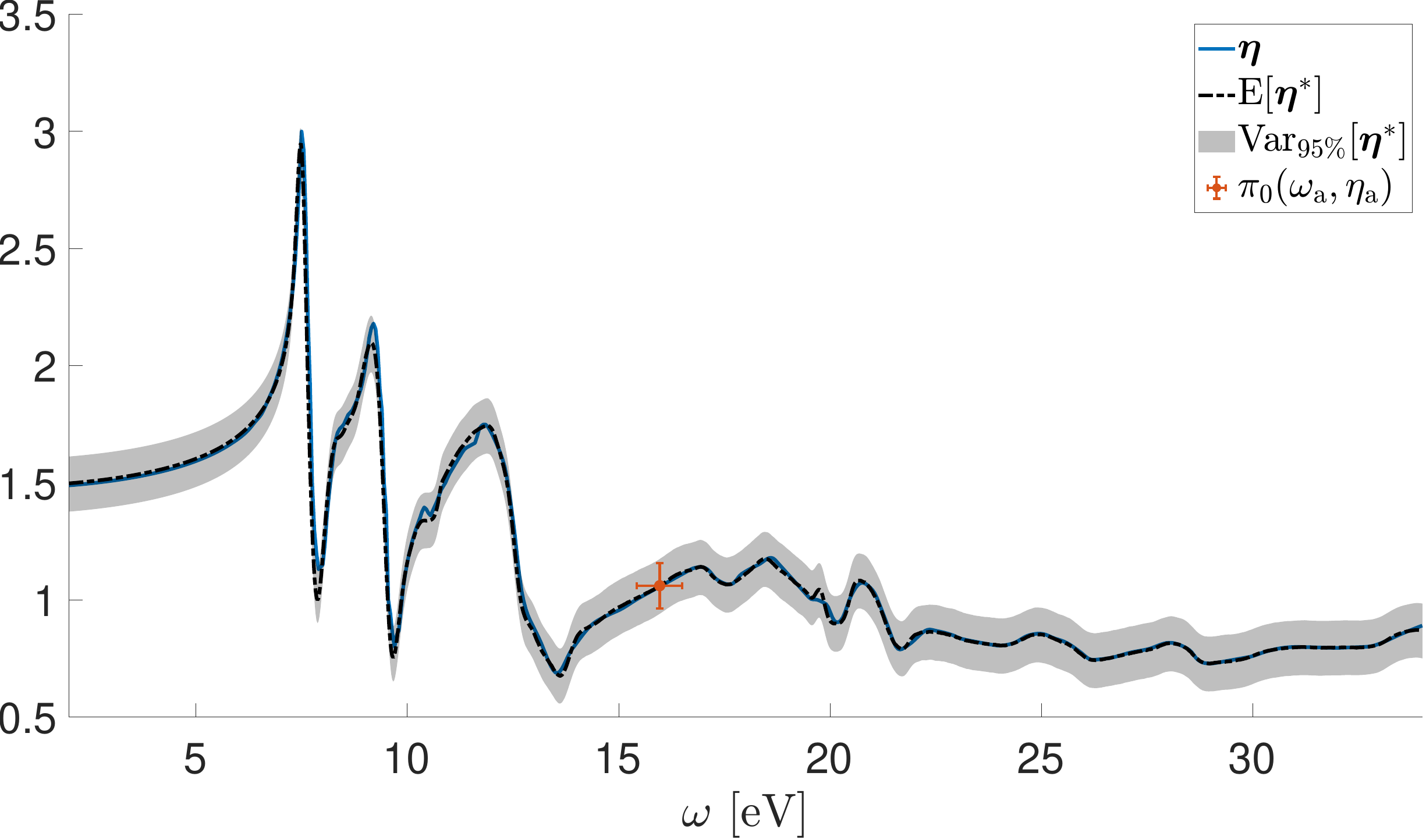}
    \includegraphics[width = \linewidth]{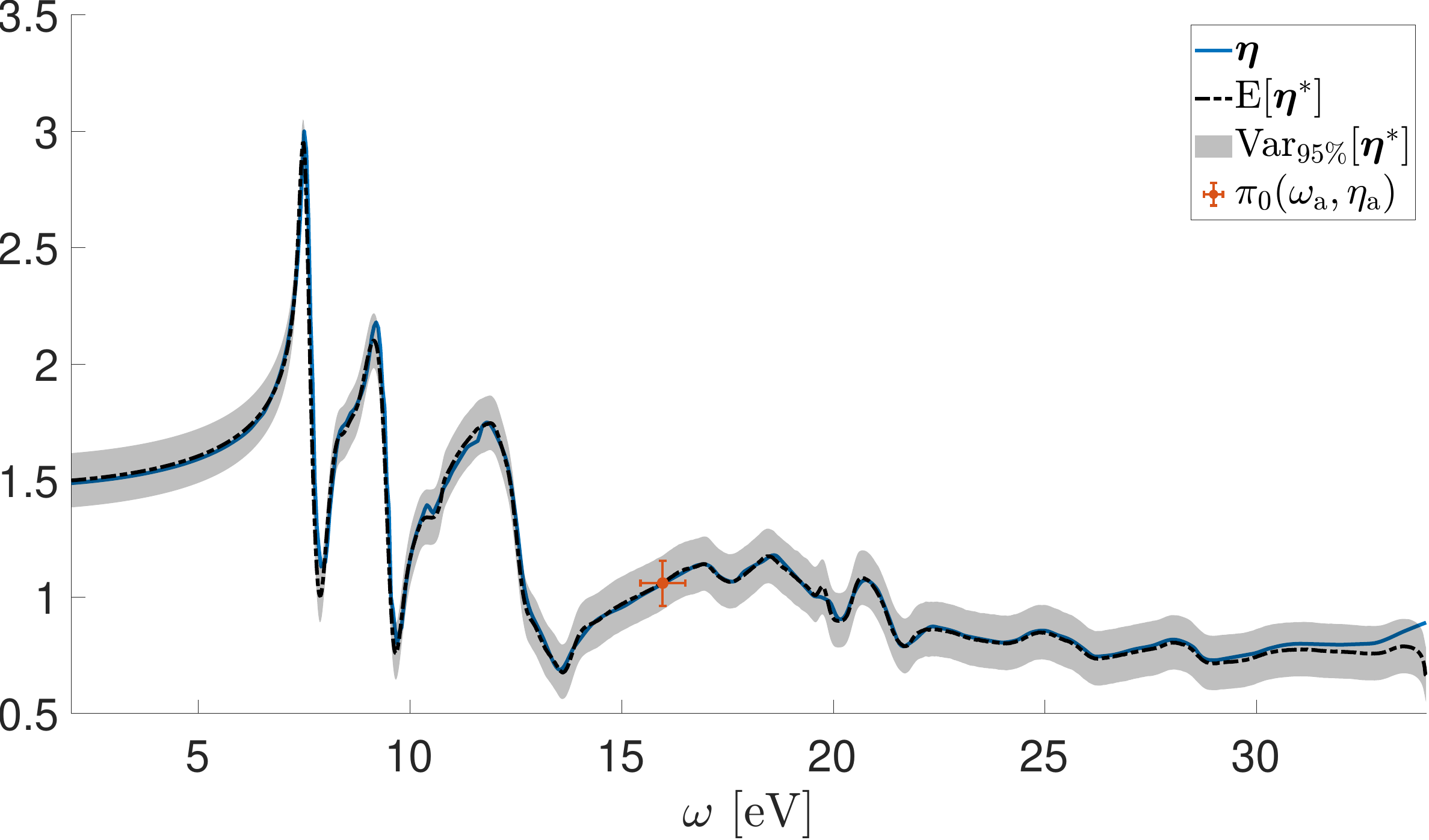}
    \caption{Results for the potassium chloride dataset. At the top, attenuation measurement data $\boldsymbol{\kappa}$, together with estimated predictive mean and 95\% marginal predictive intervals denoted by $ \rm{E} \left[ \boldsymbol{\kappa}^* \right]$ and $ \rm{Var}_{95\%} \left[ \boldsymbol{\kappa}^* \right]$, respectively. In the middle and bottom, predictive means and 95\% marginal predictive intervals $ \rm{E} \left[ \boldsymbol{\eta}^* \right]$ and $ \rm{Var}_{95\%} \left[ \boldsymbol{\eta}^* \right]$ for the real, refractive component estimated with and without extrapolation, respectively. The reference measured refractive index is denoted by $\boldsymbol{\eta}$ and the 95\% intervals of the anchor point distribution $\pi_0( \omega_\text{a}, \eta_\text{a} )$ are illustrated in red.}
    \label{im:kclResults}
\end{figure}

\begin{figure}
    \centering
    \includegraphics[width = \linewidth]{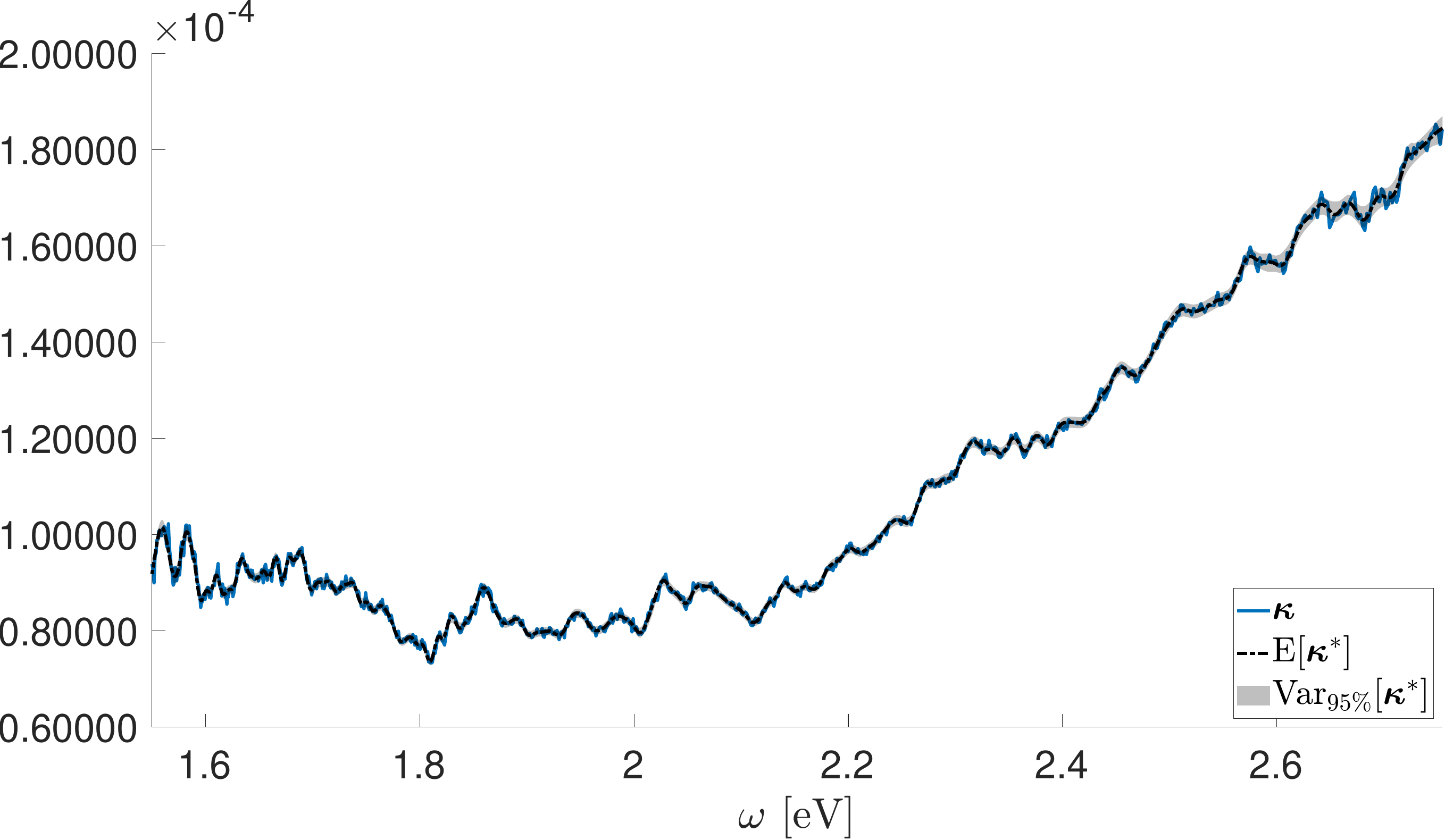}
    \includegraphics[width = \linewidth]{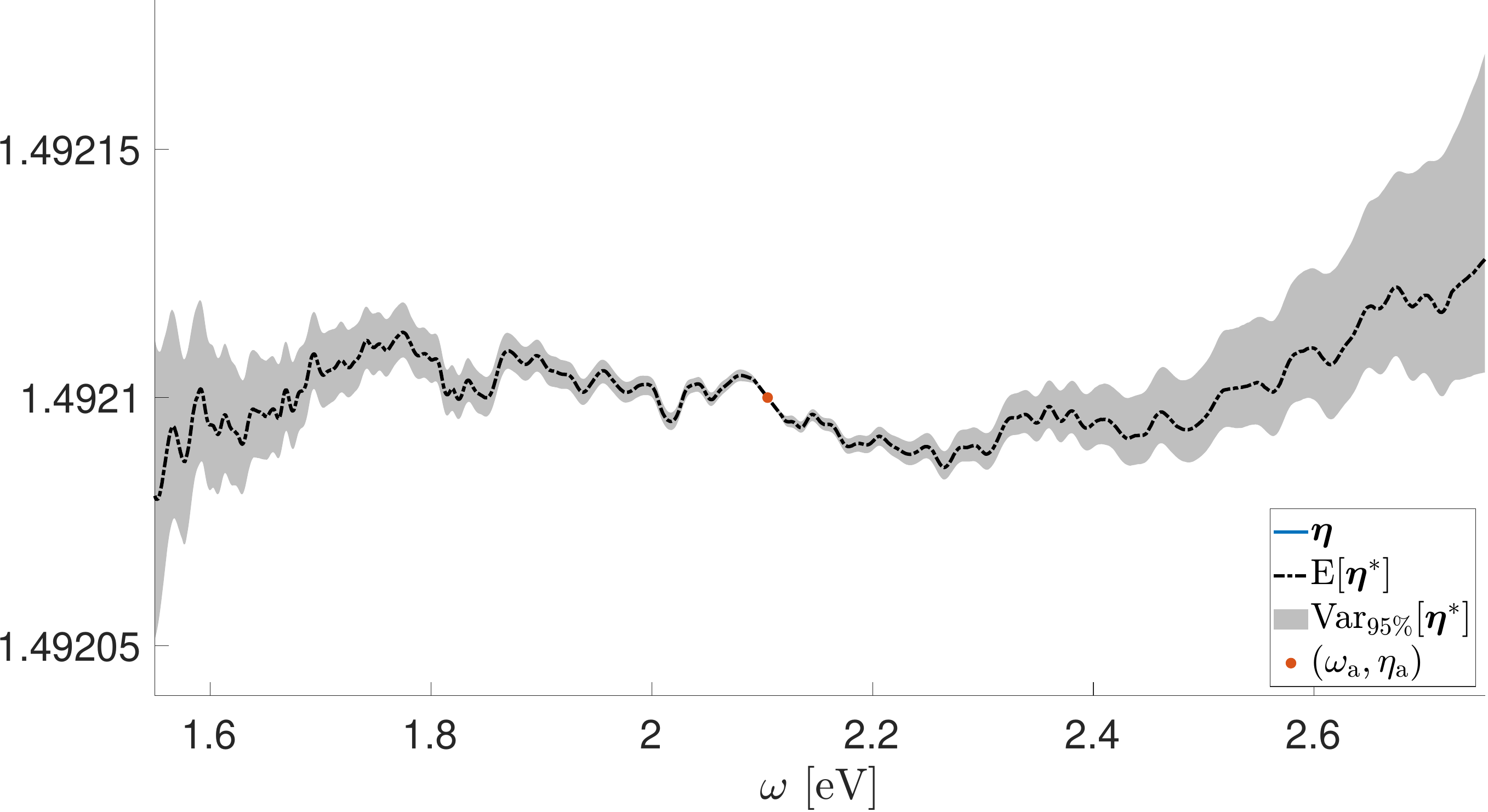}
    \caption{Results for the transparent wood dataset without background correction. At the top, attenuation measurement data $\boldsymbol{\kappa}$, together with estimated predictive mean and 95\% marginal predictive intervals denoted by $ \rm{E} \left[ \boldsymbol{\kappa}^* \right]$ and $ \rm{Var}_{95\%} \left[ \boldsymbol{\kappa}^* \right]$, respectively. At the bottom, predictive mean and 95\% marginal predictive interval $ \rm{E} \left[ \boldsymbol{\eta}^* \right]$ and $ \rm{Var}_{95\%} \left[ \boldsymbol{\eta}^* \right]$ for the real, refractive component estimated with extrapolation. The fixed anchor point $( \omega_\text{a}, \eta_\text{a} )$ is illustrated in red.}
    \label{im:twResults}
\end{figure}

For the gallium arsenide dataset, we present the estimated predictive mean and 95\% marginal confidence intervals for the imaginary and real parts of the complex refractive index in Figure \ref{im:gaasResults} together with an illustration of 95\% intervals of the anchor point distribution $\pi_0( \omega_\text{a}, \eta_\text{a})$.
Similarly, we present the predictive mean, marginal confidence intervals, and anchor point distribution intervals for the potassium chloride measurements in Figure \ref{im:kclResults}.
Finally, results for the transparent wood dataset are shown in Figure \ref{im:twResults}.

Attenuation predictive mean and marginal confidence intervals for the gallium arsenide and potassium chloride datasets closely follow the measurement as expected, due to the negligible measurement errors, as shown at the top in Figures \ref{im:gaasResults} and \ref{im:kclResults}.
Given the narrow confidence intervals, the realizations closely follow the estimated predictive mean, and, by proxy, the data points within the measurement domain.
This results in the anchor distribution $\pi_0( \omega_\text{a}, \eta_\text{a})$ defining the majority of the uncertainty of the real part $\boldsymbol{\eta}^*$ for the gallium arsenide and potassium chloride datasets.

The effect of extrapolation is most evident at the boundaries of the measurement data, particularly for the gallium arsenide dataset.
The flexible extrapolation of the measurement data results in a wide range of behavior for the extrapolated attenuation measurements, as illustrated in Figure \ref{im:examplePartitionRealization}.
This results in gradually wider estimated 95\% marginal confidence intervals towards the boundaries, most pronouncedly at $\omega \approx 6\text{ eV}$.
The necessity of the extrapolation is evident in Figures \ref{im:gaasResults} and \ref{im:kclResults}, where the numerical estimates at the measurement boundaries without extrapolation exhibit blowup-like behavior towards negative infinity for larger $\omega$.
Due to the gallium arsenide measurement data being relatively close to zero for small $\omega$, the blowup is less pronounced around $\omega = 1.5\text{ eV}$ in Figure \ref{im:gaasResults}.
For potassium chloride measurements, the measurements are zero for approximately $\omega \leq 7\text{ eV}$ and the extrapolation is only necessary for growing $\omega$.
For both datasets, the results are in good agreement with the measured real, refractive part of the complex refractive index.

The transparent wood data offers a compelling, practical example for modeling the anchor point as a probability distribution.
As the wood sample is a composite of polymethyl methacrylate (PMMA) and the actual wood substrate, together with random air voids within the substrate, the refractive index is inherently uncertain and different depending on the exact location of the composite.
However, the uncertainty of incorporating the refractive indices of PMMA (1.490), wood substrate (1.536), and, in particular, air (1.000) dwarfs the uncertainty of the small attenuation measurements as the material is transparent.
Therefore, we opt to present the results for the transparent wood sample with a fixed anchor point for visual clarity and interest.

In contrast to the gallium arsenide and potassium chloride datasets, the transparent wood dataset has noticeable measurement noise.
This results in larger predictive marginal intervals for the attenuation measurement while still overall following the trend of the data points as can be seen in Figure~\ref{im:twResults}.
Given the fixed anchor point, the uncertainty for the real part of the complex refractive index is zero at the anchor point and gradually grows towards the boundaries, highlighting the uncertainty introduced by the extrapolation.
\begin{figure}
    \centering
    \includegraphics[width = \linewidth]{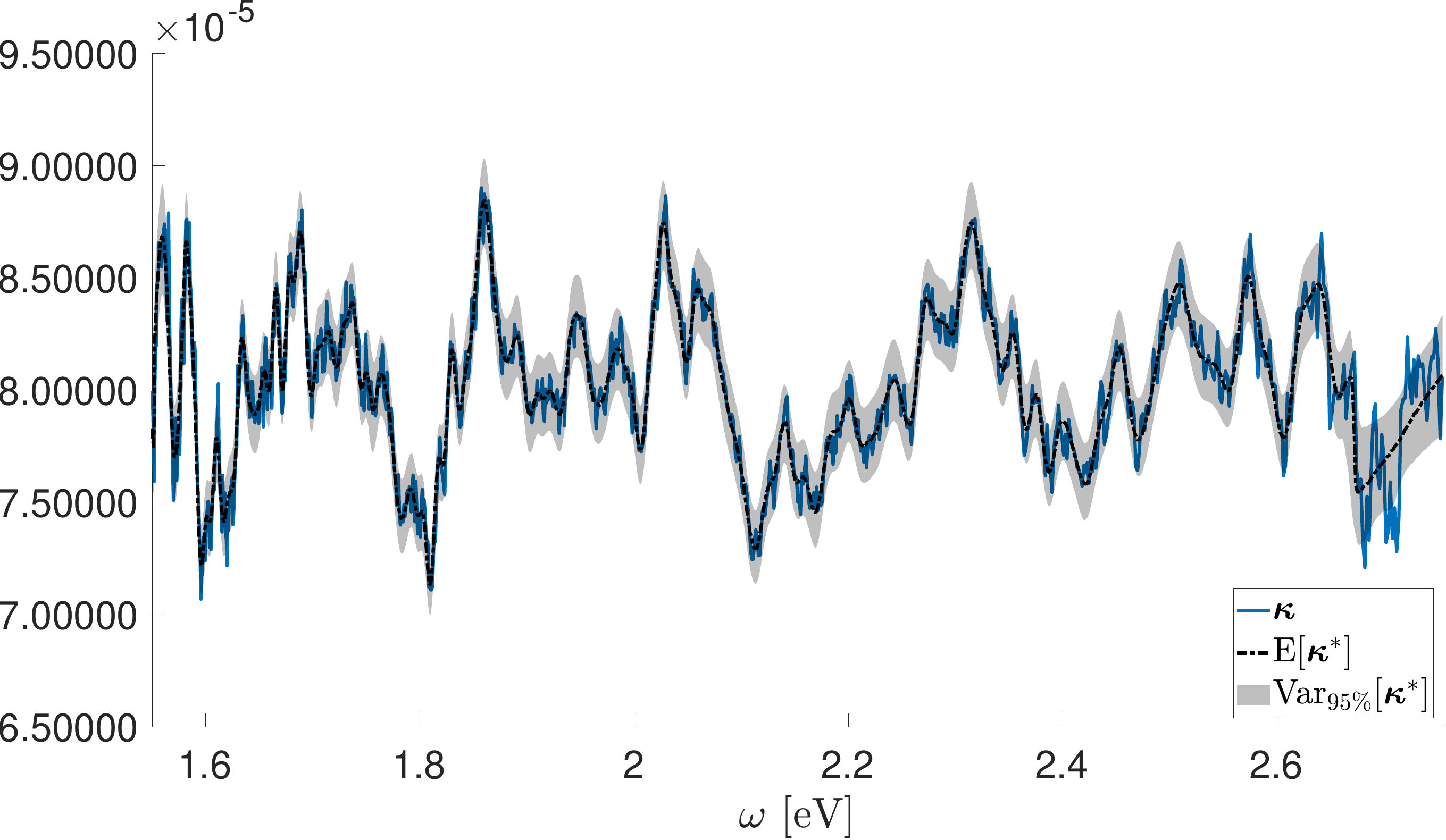}
    \includegraphics[width = \linewidth]{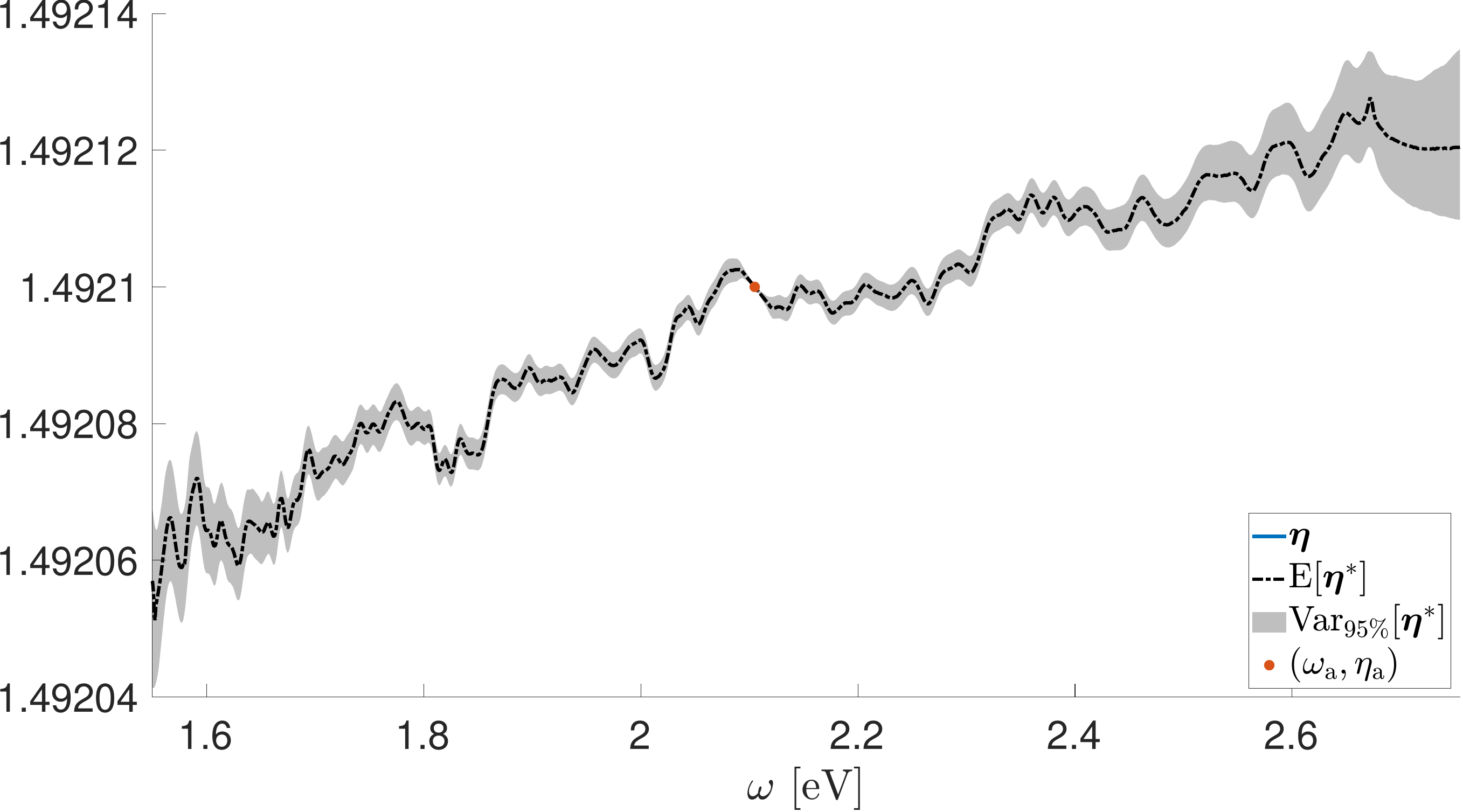}
    \caption{Results for the transparent wood dataset with background correction. At the top, attenuation measurement data $\boldsymbol{\kappa}$, together with estimated predictive mean and 95\% marginal predictive intervals denoted by $ \rm{E} \left[ \boldsymbol{\kappa}^* \right]$ and $ \rm{Var}_{95\%} \left[ \boldsymbol{\kappa}^* \right]$, respectively. At the bottom, predictive mean and 95\% marginal predictive interval $ \rm{E} \left[ \boldsymbol{\eta}^* \right]$ and $ \rm{Var}_{95\%} \left[ \boldsymbol{\eta}^* \right]$ for the real, refractive component estimated with extrapolation. The fixed anchor point $( \omega_\text{a}, \eta_\text{a} )$ is illustrated in red.}
    \label{im:twbgResults}
\end{figure}
\section{Conclusions}
\label{sec:conclusions}
We propose a statistical method for estimating the complex refractive index from attenuation measurements and a single \textit{a priori} known anchor point for the real part of the refractive index.
The model is based on mixtures of Gaussian process experts, which partition measurement data into distinct sets that are modeled using independent Gaussian processes.
The covariance function of the Gaussian processes is constructed as a linear combination of a squared exponential and a linear kernel.
The linear kernel, followed by exponentiation, encodes an exponential decay, or growth, for the model realization.
This linear combination allows for flexible extrapolation of the measurement data for statistical estimates of the Kramers-Kronig relations.
Furthermore, the extensive Bayesian and statistical treatment of model quantities and the anchor points yields a probability distribution for the complex refractive index, allowing for uncertainty quantification or further propagation of uncertainty in subsequent analyses.
Importantly, the method is completely automatic, requiring no manual selection of extrapolation points or explicit extrapolation model selection.

We apply the method to three experimental attenuation datasets of gallium arsenide, potassium chloride, and a transparent wood sample.
For the gallium arsenide and potassium chloride samples, we also have access to measurements of the real part of the complex refractive index, which are used to validate the computational approach.
The results show a clear improvement in the complex refractive index estimate, particularly at the boundaries of the measurement region, compared to the statistical model applied without extrapolation.

For future work, it would be interesting to compare the behavior of the Hilbert-transform-constrained kernel\cite{Ciucci:2020} to the more standard squared exponential kernel used here to make the approach more physics-informed.
Integrating additional physical constraints, such as sum rules or known asymptotic behavior of optical constants, directly into the prior structure of the mixture model could offer a useful way to combine data‑driven flexibility with stronger physics‑based regularization.
Naturally, applying the methodology to further measurement datasets, particularly ones with significant measurement noise, would be an interesting practical avenue of future research.
Furthermore, the methodology could be extended to multi-angle reflectometry or spectroscopic ellipsometry data, where the joint estimation of $n$ and $\kappa$ is more involved.
Incorporating non-Gaussian noise distributions would also provide insights into its robustness in real-world experimental conditions.
However, this would require significant methodological developments as existing inference methods, such as the nested sequential Monte Carlo sampler used, rely explicitly on analytical properties of Gaussian processes.

\section*{Author contributions}
T.M.: Conceptualization, Methodology, Software, Validation, Formal analysis, Writing - Original Draft, Visualization
H.C.: Investigation, Resources, Writing - Original Draft
E.M.V.: Investigation, Resources, Writing - Original Draft

\section*{Conflicts of interest}
There are no conflicts to declare.
\section*{Data availability}
The software for the statistical algorithms described in this study can be found at \href{https://doi.org/10.5281/zenodo.19235589}{Zenodo} with DOI: 10.5281/zenodo.19235589. The version of the code employed for this study is the version corresponding to the DOI.
\section*{Acknowledgements}
The authors thank the HORIZON AI-TRANSPWOOD (AI-Driven Multiscale Methodology to Develop Transparent Wood as Sustainable Functional Material) project, Grant no. 101138191, co-funded by the European Union. Views and opinions expressed are however those of the author(s) only and do not necessarily reflect those of the European Union or HaDEA. Neither the European Union nor the granting authority can be held responsible for them.

%%%END OF MAIN TEXT%%%


\balance


%%%REFERENCES%%%
\bibliography{ref.bib} %You need to replace "rsc" on this line with the name of your .bib file
\bibliographystyle{rsc} %the RSC's .bst file

\end{document}